\documentstyle[12pt,epsf]{article}
\newcommand{\abstracts}[1]{{
\centering{\begin{minipage}{12.2truecm}
\normalsize\baselineskip=15pt
\centerline{\footnotesize ABSTRACT}\vspace*{0.3cm}
\parindent=20pt #1
\end{minipage}}\par}}
\newcommand{\dual}{\mbox{}^{\ast}}
\newcommand{\dd}{\mbox{\rm d}}

\newcommand{\beqn}{\begin{eqnarray}}
\newcommand{\eeqn}{\end{eqnarray}}
\newcommand{\eq}[1]{(\ref{#1})}
\newcommand{\beq}{\begin{equation}}
\newcommand{\eeq}{\end{equation}}

\newcommand{\cU}{{\cal U}}

\begin{document}
~\vspace{-1.5cm}
\begin{flushright}
{\large KANAZAWA-97-05}
\end{flushright}
\vspace{1.5cm}

\begin{center}

{\baselineskip=16pt
{\Large \bf Mass gap, Abelian Dominance and Vortex Dynamics
in $SU(2)$ Spin Model}\\

\vspace{1cm}

{\large O.A.~Borisenko
\footnote{email: oleg@ap3.gluk.apc.org}$^{\! ,a}$ ,
M.N.~Chernodub\footnote{email:
chernodub@vxitep.itep.ru}$^{\! ,b,c}$ and F.V.~Gubarev\footnote{email:
gubarev@vxitep.itep.ru}$^{\! ,b}$}\\

\vspace{.5cm}
{ \it

$^a$ N.N.Bogolyubov Institute for Theoretical Physics, National\\ Academy
of Sciences of Ukraine, 252143 Kiev, Ukraine

\vspace{0.3cm}

$^b$ ITEP, B.Cheremushkinskaya 25, Moscow, 117259, Russia

\vspace{0.3cm}

$^c$ Department of Physics, Kanazawa University,\\
Kanazawa 920-11, Japan

}
}
\end{center}

\vspace{1cm}

\abstracts{
We discuss a new approach to the investigation of the nature of the
mass gap in spin systems with continuous global symmetries which is
much analogous to the method of abelian projection in the gauge
theories. We suggest that the abelian degrees of freedom, in
particular, abelian vortices are responsible for the mass gap
generation phenomena in the non-abelian spin systems. To check our
hypothesis we study numerically the three--dimensional $SU(2)$ spin
model in the Maximal Abelian projection. We find that the abelian
mass gap in the projected theory coincides with the full non-abelian
mass gap within numerical errors.  The study of the percolation
properties of the abelian vortex trajectories shows that the phase
transition and the mass gap generation in the $3D$ $SU(2)$ spin model
are driven by the abelian vortex condensation.  }

\newpage

\setcounter{footnote}{0}
\renewcommand{\thefootnote}{\alph{footnote}}

\section{Introduction}
\baselineskip=14pt

This paper proposes a new approach to the investigation of the mass
gap nature in the spin systems with nonabelian symmetry group.  While
this question is well studied in the context of $O(2)$ spin system,
it is open for non-abelian spin systems. The situation much resembles
the problem with the string tension in pure gauge theory where the
monopoles are known to be responsible for the confinement in the
strong coupling lattice $U(1)$ theory \cite{BaMyKo77} but the nature
of the nonzero string tension in a nonabelian theory at
arbitrary couplings
is still unclear. The main question of all studies devoted to these
problems is what are relevant configurations which lead to nonzero
string tension in gauge theories or mass gap in spin systems,
correspondingly.

During last decade there was some progress in understanding the
confining forces in the $SU(N)$ gauge theories. In particular, this
progress was done in the framework of partially fixed gauges, for
example, within abelian projection approach~\cite{tH81}.  This
approach is based on the partial gauge fixing of the $SU(N)$ gauge
degrees of freedom to an ``abelian gauge'', such that the maximal
torus group ${[U(1)]^{N-1}}$ remains unfixed. After the partial gauge
fixing, abelian monopoles appear due to compactness of the residual
abelian gauge group. According to the dual superconductor picture
\cite{tH76Ma76}, the condensation of such monopoles may explain the
colour confinement
in gluodynamics: if the monopoles are condensed
then the confining string is formed between the quark-antiquark pair
due to the dual Meissner effect. This picture has been confirmed in
various numerical simulations of lattice gluodynamics (see, for
example, the reviews~\cite{Po96,Su93}).

The mass gap has also been intensively investigated in the context of
spin models. Berezinskii, Kosterlitz and Thouless \cite{BKT} showed
that the condensation of vortices in $2D$ $XY$ model leads to the
spin-spin correlations which fall down exponentially as distance
between spins increases. Thus the vortex condensation gives rise to
the nonzero mass gap in the relevant region of the coupling constant
\cite{BKT}. Up to now it is not much known, however about the nature
of the mass gap in nonabelian spin systems (see, \cite{NatureMassGap}
and references therein on the general review of the subject). The
usual methods for the studying of the mass gap in the spin models
(high and low temperature expansions, Monte-Carlo simulations) do not
deal much with the nature of the mass gap. These methods rather
concentrate on other problems like behaviour of the mass gap in the
finite volume and in the perturbation theory, comparison of the mass
gap which is calculated in Monte-Carlo simulations with the Bethe
ansatz prediction, {\it etc.} (see, for instance, \cite{MassGap}).
Even ``exact'' mass gap (calculated in 2D $O(N)$ models under
assumptions of validity of the Bethe ansatz and the standard
perturbation theory) tells us exactly nothing about relevant
configurations of spins contributing to the mass gap \cite{BetheAnz}.

Meantime, it would be very desirable and suggestive to have a deeper
insight into the nature and properties of the mass gap in nonabelian
spin models. We are interested in a possible generalization of the
vortex mechanism of the mass gap generation in the $XY$ model to the
nonabelian spin systems. One can also elaborate mechanisms of the
mass gap generation, which are analogous to the confinement
mechanisms in gauge theories. It seems to be natural then to apply
known methods of the investigation of abelian degrees of freedom in
gauge systems to the spin systems. In this paper we propose
to extend the method of abelian gauge fixing to the spin systems in
order to find spin configurations which are relevant for the mass gap
generation. The Maximal Abelian (MaA) projection \cite{KrLaScWi87}
is of a special interest for us in this respect since this projection
is a good framework to display the dynamics of abelian degrees of
freedom of the gauge theory.

In Section 2 we define the Maximal Abelian projection for $SU(2)$
spin model and we show that after the abelian projection the $SU(2)$
spin model becomes ${[O(2)]}^2$ spin model with non-local interaction
between spins. This abelian system possesses two types of the $O(2)$
vortices. We argue that these vortices as well as the corresponding
abelian spins may have different relevance to the dynamics of the
$SU(2)$ spin system in the MaA projection. We also stress the
difference between projections in gauge and spin models. The present
publication is devoted to the study of most relevant type of the
abelian degrees of freedom. More detailed investigation of the
properties of abelian system as well as the discussion of the $SU(N)$
spin models will be presented elsewhere \cite{InPreparation}.

In Section~3 we study numerically the abelian mass
gap which is obtained from abelian spin--spin correlators in
the MaA projection and compare it with the $SU(2)$ mass gap. We find
that $SU(2)$ mass gap coincides with its abelian counterpart within
numerical errors. This result is similar to the effect of
``abelian dominance'' in lattice gluodynamics~\cite{SuYo90,Ez82}.
The investigation of the vortex percolation properties
shows that the abelian vortices are condensed in the massive phase and
they are dilute in the massless phase.

We conclude that i) the abelian degrees of freedom
in the MaA gauge
are relevant degrees of freedom for the mass gap
generation in $SU(2)$ spin model;
ii) the mechanisms of the mass gap generation in $3D$ $SU(2)$ spin
system in the MaA projection is similar to that in $3D$ $XY$
model~\cite{KoShWi86,PoPoYu91,AH93-94}: the
abelian vortex condensation leads to the mass gap generation.

\section{Maximal Abelian Projection for $SU(2)$ Spins}

We consider the $SU(2)$ spin model with the action

\beqn
S = - \frac{\beta}{2} \sum_x \sum^D_{\mu=1} {\rm Tr} \, U_x \,
    U^+_{x+\mu} \,, \label{actionSU2}
\eeqn
where $U_x$ are the $SU(2)$ spin fields, $\beta = \frac{2}{g^2}$ and
$D$ is the dimension. The action \eq{actionSU2} is invariant under
$SU_L(2) \times SU_R(2)$ global transformations:

\beqn
U_x  \to U^{(\Omega)}_x =
\Omega^+_L \, U_x \, \Omega_R\,,\quad \Omega^+_{L,R} \in SU(2)
\,. \label{SU2Transf}
\eeqn

In analogy with the abelian projection in the non--abelian gauge
theories \cite{tH81}, we define the abelian projection in the
non-abelian spin systems as a gauge fixing of the non-abelian
symmetry group up to its maximal torus subgroup.

Let us recall the definition of the MaA projection in $SU(2)$ lattice
gluodynamics~\cite{KrLaScWi87}. The gauge fixing condition is defined as
the maximization of the functional

\beqn
  R_g [\cU] & = & \sum_{x,\mu} {\rm Tr}
  \Bigl( \cU_{x,\mu} \sigma^3 \cU^+_{x,\mu} \sigma^3 \Bigr)\,,
  \label{Rgauge}
\eeqn
over all $SU(2)$ gauge transformations $\cU_{x,\mu} \to \Omega^+_x
\cU_{x,\mu}\Omega_{x+ \hat \mu} $,

\beqn
  \max_{\{ \Omega \}} R_{gauge}[\cU^{(\Omega)}]\,,
  \label{Rmax}
\eeqn
here $\cU_{x,\mu}$ are the $SU(2)$ gauge link fields
and $\sigma^a$ are the Pauli matrices.

The functional $R_g$ is invariant under $U(1)$ gauge
transformations ${\tilde \Omega}_x = e^{i \sigma_3 \, \alpha_x}$,
$\alpha \in [0,2 \pi)$, and therefore it fixes the $SU(2)$ gauge
group up to the diagonal $U(1)$ subgroup. The residual abelian group is
compact, and the abelian projected theory contains the abelian
monopoles. The
numerical simulations of the $SU(2)$ gluodynamics in the Maximal Abelian
projection shows that the abelian degrees of freedom, in particular
the abelian monopoles are responsible for the confinement
phenomenon \cite{Po96,Su93}.
There are a lot of the abelian projections of the $SU(2)$
gluodynamics but the abelian degrees of freedom in the MaA projection
seem to be the most relevant to infrared properties of this gauge
theory. Therefore we may expect that the abelian degrees of freedom
in the spin system in the MaA projection are responsible for the
infrared structure of the spin model.

We define the Maximal Abelian projection for the $SU(2)$ spin theory
by the following maximization condition:

\beqn
  \max_{\{ \Omega \}} R_s [U^{(\Omega)}]\,,
  \label{maxR}
\eeqn
where the maximized functional is the analog of functional \eq{Rgauge}:

\beqn
  R_s [U] & = & \sum_x {\rm Tr} \Bigl( U_x \sigma^3 U^+_x
  \sigma^3 \Bigr)\,,
  \label{R}
\eeqn

The functional $R_s [U]$ is invariant under $U_L(1) \times U_R (1)$
global transformations:

\beqn
 U_x & \to & U^{({\tilde \Omega})}_x =
 {\tilde \Omega}^+_L U_x {\tilde \Omega}_R\,; \quad
 {\tilde \Omega_{L,R}} = e^{i \sigma^3 \omega_{L,R}}\,, \quad
 \omega_{L,R} \in [0, 2 \pi)\,.
  \label{Ug}
\eeqn
Due to the invariance \eq{Ug}, the condition \eq{maxR}
fixes the $SU_L(2) \times SU_R (2)$
global symmetry group up to $U_L(1) \times U_R (1) \sim O_L (2)
\times O_R (2)$ global group.

The important difference between the abelian gauge fixing for the
spin and for the gauge systems is that in the spin system we fix the
{\it global} symmetry. Thus the number of the gauge conditions does
not depend on the lattice volume while the symmetry group in the
gauge system is local and the number of the gauge conditions grows
proportionally to the volume.  Therefore the projection in gauge
system fixes each link matrix to be as close to diagonal as possible,
while the projection in the spin system forces an average value of
the spin to be diagonal.  Therefore, the projection for the spin
systems is somehow simpler to handle than for gauge theories.

Let us parametrize the $SU(2)$ spin field $U$ in the standard way:
$U^{11}_x = \cos \varphi_x e^{i\theta_x}; $
$\ U^{12}_x = \sin
\varphi_x e^{i\chi_x}; \ U^{22}_x = U^{11, *}_x; \
U^{21}_x = - U^{12,*}_x; \ 0 \le \varphi \le \pi/2, \ 0 <
\theta,\chi \le 2 \pi$. Under the $O_L(2) \times O_R (2)$ transformation
(\ref{Ug}), components of the field $U$ transform as

\beqn
 \theta_x \to \theta^\prime_x
 = \theta_x + \omega_d \quad {\rm mod}\, 2 \pi\,,\qquad
 \chi_x \to \chi^\prime_x & =
 \chi_x + \omega_o \quad {\rm mod}\, 2 \pi\,,
\eeqn
where $\omega_{d,o} = - \omega_L \mp \omega_R $. It is
convenient to decompose the residual symmetry group,
$O_L (2) \times O_R (2) \sim O_d (2) \times O_o (2)$. The diagonal
(off-diagonal) component $\theta$ ($\chi$) of the $SU(2)$ spin $U$
transforms as a spin variable with respect to the $O_d(2)$ ($O_o(2)$)
symmetry group.

It is useful to consider the one-link spin action $S_l$
in terms of the angles $\varphi$, $\theta$, $\chi$:

\beqn
S_{l_{x,\mu}}
= - \beta \Bigl[\cos\varphi_x \cos\varphi_{x+\hat\mu}
\cos(\theta_x - \theta_{x+\hat\mu})
+ \sin\varphi_x \sin\varphi_{x+\hat\mu}
\cos(\chi_x - \chi_{x+\hat\mu})
\Bigr]\,.
\label{SU2Act}
\eeqn
The action consists of two parts which correspond to the
self--interaction of the spins $\theta$ and $\chi$, respectively. The
$SU(2)$ component $\varphi$ does not behave as a spin field and its
role is to provide the interaction between the $\theta$ and $\chi$
spins. Thus, the $SU(2)$ spin model in the abelian projection reduces
to two interacting copies of the $XY$ model with the fluctuating
couplings due to the dynamics of the field $\varphi$.

Two types of the abelian vortices appear in the abelian projected
$SU(2)$ spin model due to the compactness of the residual abelian
group. They correspond to the diagonal ($\theta$) and to the
off-diagonal ($\chi$) abelian spins. We call these vortices as
``diagonal'' and ``off-diagonal'' vortices, respectively.

We expect that in the MaA projection the diagonal
vortices may be more dynamically important then
the off-diagonal vortices. The reason for this expectation is simple.
In our representation for $SU(2)$ spin field
the maximizing functional \eq{R} has the form:
$R_{spin} = 4 \sum_x \cos^2 \varphi_x + const$.
Therefore, in the MaA projection the effective coupling constant
in front of the action for the $\theta$
 spins is maximized while
the effective self--coupling for $\chi$ spins
is minimized. Thus we may expect that the diagonal vortices
may be more relevant to the dynamics of the
system than the off-diagonal vortices\footnote{There exists similar
situation in the lattice gluodynamics in the
MaA projection: the action for the
diagonal variables of the $SU(2)$ gauge fields is larger
then the action for the non-diagonal variables \cite{ChPoVe95}.
The numerical simulations show \cite{ChPoVe95} that
topological defects in the diagonal fields, monopoles,
are related to the dynamics of the gauge theory,
while off--diagonal defects, minopoles, decouple from the
dynamics of the gauge system.}.

One should note that both types of the abelian vortices as well as any
other ${[O(2)]}^2$ invariant quantities in the MaA projection can be
easily defined in the $SU(2)$ invariant way \cite{InPreparation}.
Therefore the abelian observables which are discussed in our work
cannot be gauge artifacts.

\section{Results of Numerical Simulations}

In the present Section we study the abelian mass gap for the diagonal
spins and the behaviour of the diagonal vortices in the three
dimensional $SU(2)$ spin model on the lattices $L^2 \times L_z$,
$L=16$, $L_z=16,32$ with periodic boundary conditions.

We calculate the non-abelian mass gap from the plane-plane
correlator:

\beqn
<{\rm Tr}(
{\tilde U}(0)\cdot {\tilde U}^+(z))> = const.\,
\Bigl( e^{ - \mu_{SU(2)} \cdot z } +
e^{ - \mu_{SU(2)} \cdot (L_z - z) }
 \Bigr)
+ \dots\,,\label{full}
\eeqn
where ${\tilde  U}(z) = L^{-1} \, \sum^{L}_{x,y=1} U(x,y,z)$ is the
zero-momentum spin operator in the $x-y$ plane and terms vanishing in
the limit $L_z\to \infty$ are denoted by dots. Mass gap is calculated
by fitting the zero--momentum spin--spin correlator by the
leading term in eq.\eq{full}.

The abelian mass gap can be calculated similarly. We use the
zero-momentum {\it abelian} spin operator ${\tilde  U}_{ab} (z)=
L^{-1} \, \sum^{L}_{x,y=1} \exp\{i
\, \theta(x,y,z)\}$ and measure
the correlator:

\beqn
{<{\tilde U}_{ab}(0)\cdot {\tilde U}^*_{ab}(z)>}_{MaA} = const.\,
\Bigl( e^{ - \mu_{O(2)} \cdot z } + e^{ - \mu_{O(2)}
\cdot (L_z - z) } \Bigr)
+ \dots\,, \label{abelian}
\eeqn
where the quantum average is taken in the MaA gauge (\ref{maxR},\ref{R}).
Zero momentum 
correlator \eq{abelian} can be defined~\cite{InPreparation} 
as a quantum average of some non-local $SU(2)$--invariant operator;
this average can be calculated without gauge--fixing.
This shows that in a general gauge
the zero--momentum operator which corresponds to the
abelian spin--anti-spin in the MaA gauge
and the zero--momentum non--abelian spin--anti-spin operator 
are related to each other in a non--local way.

In numerical simulations we use U.Wolff's cluster
algorithm~\cite{Wolff}, which is known to work perfectly well in
the $O(N)$ spin models. In particular, this algorithm has no problems
with
critical slowing down. In order to thermalize our fields at each
$\beta$ value we performed a number of thermalization sweeps,
which is much greater then measured auto-correlation time.

The abelian and the $SU(2)$ mass gaps {\it vs.} $\beta$ are shown on
Fig.~1. The plotted mass gap is very small in the massless phase
($\beta > \beta_c = 0.935\dots$) but it is non zero due to the
finite-volume effects and fitting technique.  The abelian mass gap
coincides with the $SU(2)$ mass gap within numerical errors.  This
result is similar to the already known numerical result in the
lattice gluodynamics:  the abelian string tension (derived from
abelian Wilson loops) almost coincides with the full $SU(2)$ string
tension \cite{SuYo90,BaBoMuSc96}. This effect was first numerically
observed in gluodynamics in Ref.~\cite{SuYo90}, and it is called the
``abelian dominance'' \cite{SuYo90,Ez82}. In the present paper we
find the abelian dominance in the $SU(2)$ spin system: the
non--abelian mass gap generation is dominated by the abelian
degrees of freedom.

Due to the abelian dominance the abelian topological defects
(abelian vortices) may play a role in the mass gap generation. The
vortex trajectory $\dual j$ is constructed from the spin variables
$\theta$ by the formula \cite{BaMyKo77,Sa78}:  $\dual j = {(2
\pi)}^{-1} \dual \dd [\dd \theta]$, where the square brackets stand
for ``modulo $2 \pi$'' and the operator ``d'' is the lattice
derivative\footnote{We use the formalism of the differential
forms on the lattice \cite{BeJo82}.}.
In the three dimensions the vortex
trajectories are loops which are closed due to the property
$\dd^2=0$.

In order to have a feeling on the behaviour of the vortices we
visualized the vortex trajectories in the different phases of the
spin system. The examples of the typical diagonal vortex
configurations in the massive ($\beta=0.8$) and the massless
($\beta=1.1$) phases are shown on Fig.~2(a) and Fig.~2(b),
respectively.

It is clearly seen from Figs.~2(a,
b) that the diagonal vortices form
large clusters in the massive phase and they are dilute in the
massless phase. The similar properties of the vortex loops have been
found~\cite{PoPoYu91,AH93-94} in the $3D$ $XY$ model where the mass
gap generation is known to be due to the vortex
condensation~\cite{KoShWi86,PoPoYu91,AH93-94}.

One of the quantitative characteristics of the vortex ensemble
is the so-called percolation probability\footnote{This formula is
written for the infinite volume lattice. The finite volume definition
can be found in Refs.~\cite{PoPoYu91,AH93-94}.}
\cite{PoPoYu91} $C$:

\beqn
C = \lim_{r \to \infty}
{\left(
\sum\limits_{x,y,i}
\delta_{x\in \dual j_i} \, \delta_{y\in \dual j_i} \cdot
\delta\Bigl(|x-y|-r\Bigr)
\right)} \cdot {\left(
\sum\limits_{x,y} \delta\Bigl(|x-y|-r\Bigr)
\right)}^{-1}\,,
\eeqn
where the summation is going over all the vortex trajectories
$j_i$, and over all the points $x$, $y$ of the lattice.  The
quantity $C$ has a meaning of the probability for two infinitely
separated points of the lattice to be connected by a vortex
trajectory. If this probability is not zero then the infinite long vortex
loops exit in the vacuum of the spin system. The last fact implies the 
dominance of the entropy of the vortex trajectories over the 
vortex energy which means the existence of the vortex condensate.
Therefore, if the vortices are condensed this probability is
non-zero, and if the vortex condensate does not exist the quantity
$C$ must be zero \cite{PoPoYu91}.

The percolation probability $C$ for the diagonal vortex trajectories
is shown on Fig.~3. The percolation probability
is not zero in the massive phase, $\beta < \beta_c$, thus in this
phase the vortices are condensed. After the phase transition the
percolation probability is getting smaller and it vanishes right after the
phase transition point $\beta_c$. 
Therefore in the massless phase the vortices
are not condensed. Our result shows that the phase
transition in the $SU(2)$ spin model is driven by the abelian vortex
condensation.

\section*{Conclusions and Acknowledgments}

Our results show that the mass gap generation in $3D$ $SU(2)$ spin
system can be described through the abelian degrees of freedom in the
Maximal Abelian gauge. We found the abelian dominance for the mass
gap generation: the non-abelian mass gap coincides within numerical
errors with the abelian mass gap. The mechanism in the non-abelian
spin model is similar to that in $3D$ $XY$ model:  the abelian
vortex condensation leads to the mass gap generation.

In our next publication \cite{InPreparation} we are going to
investigate in detail the properties of both types of vortices and to
develop the projection method for $2D$ $SU(N)$ spin systems.

Authors are grateful to D.A.~Ozerov and M.I.~Polikarpov for fruitful
discussions. M.Ch. acknowledges the kind hospitality of the
Theoretical Department of the Kanazawa University where this work was
finished. M.Ch. and F.G. were supported by the JSPS Program on
Japan -- FSU scientists collaboration, by the Grants INTAS-94-0840,
INTAS--RFBR-95-0681,
and by the grant number 93-02-03609, financed by
the Russian Foundation for the Fundamental Sciences.

\newpage

\begin{figure}[htb]
\vspace{4cm}
\centerline{\epsfxsize=.9\textwidth\epsfbox{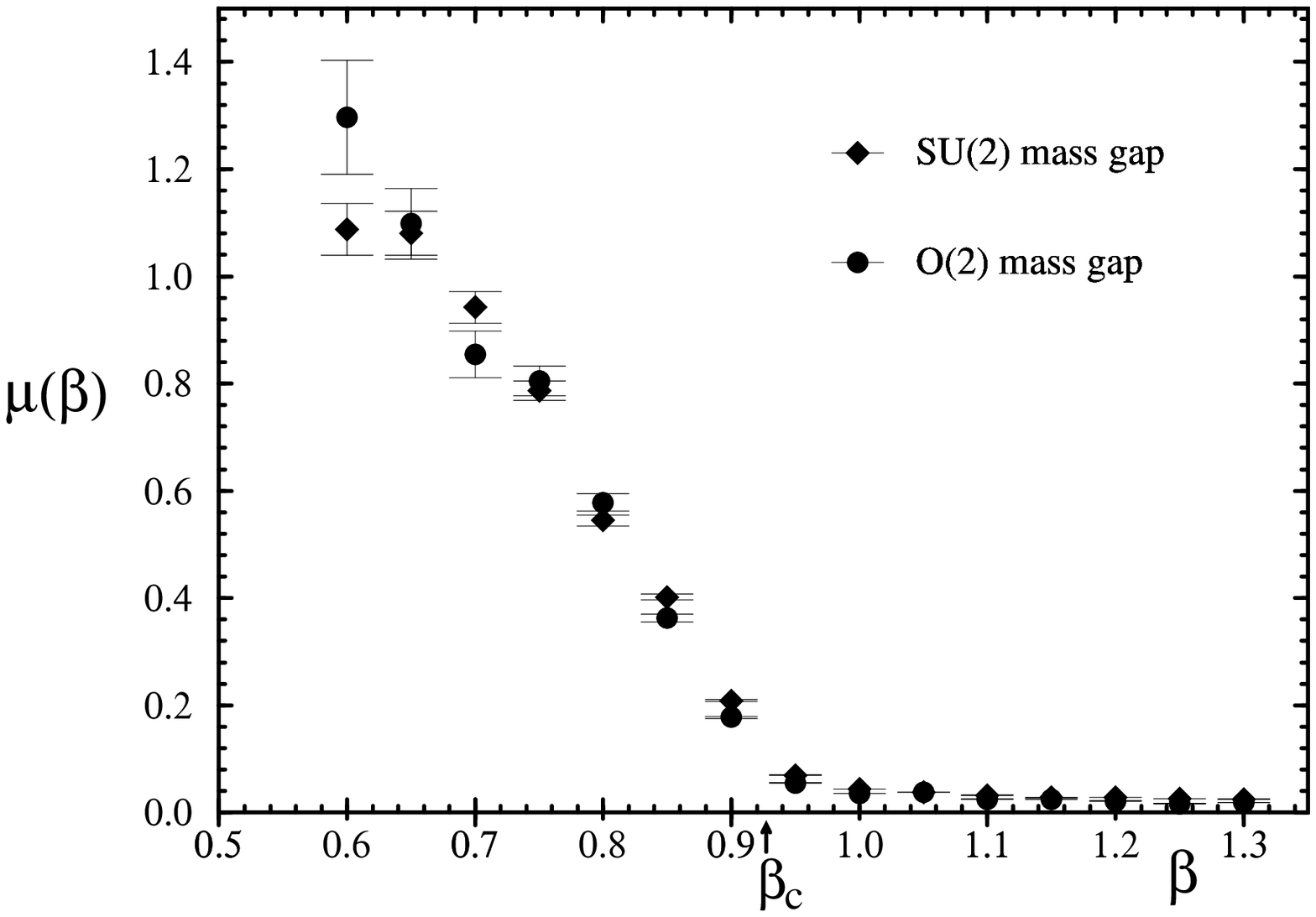}} 
\vspace{0.1cm}
\centerline{Fig.~1: The abelian and $SU(2)$ mass gaps {\it vs.}
$\beta$, the lattice is $16^2\cdot 32$.}
\end{figure}

\newpage

\begin{figure}[htb]
\vspace{-2cm}
\centerline{\epsfxsize=.60\textwidth\epsfbox{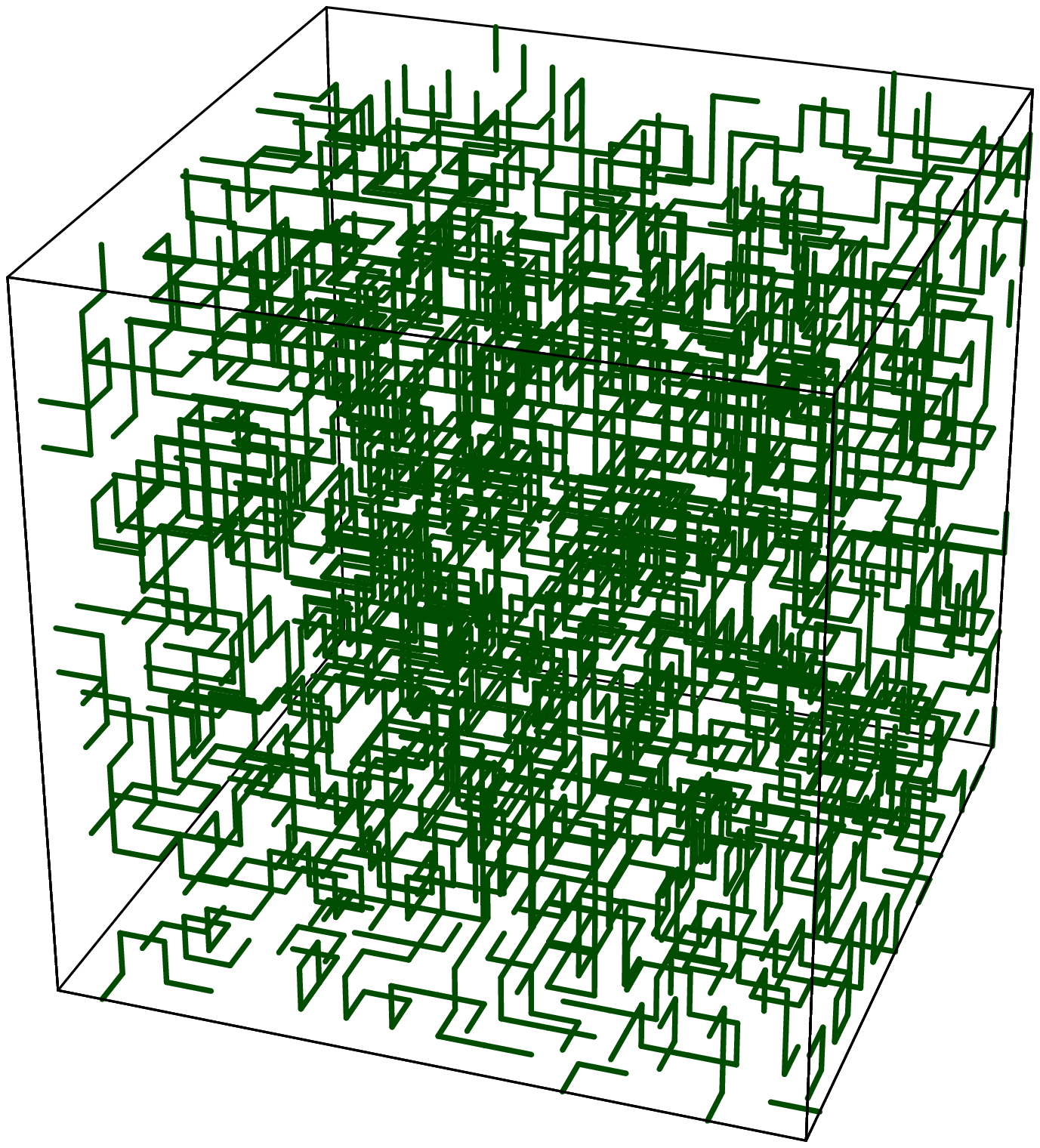}} 
\vspace{0.1cm}
\centerline{(a)}
\centerline{\epsfxsize=.60\textwidth\epsfbox{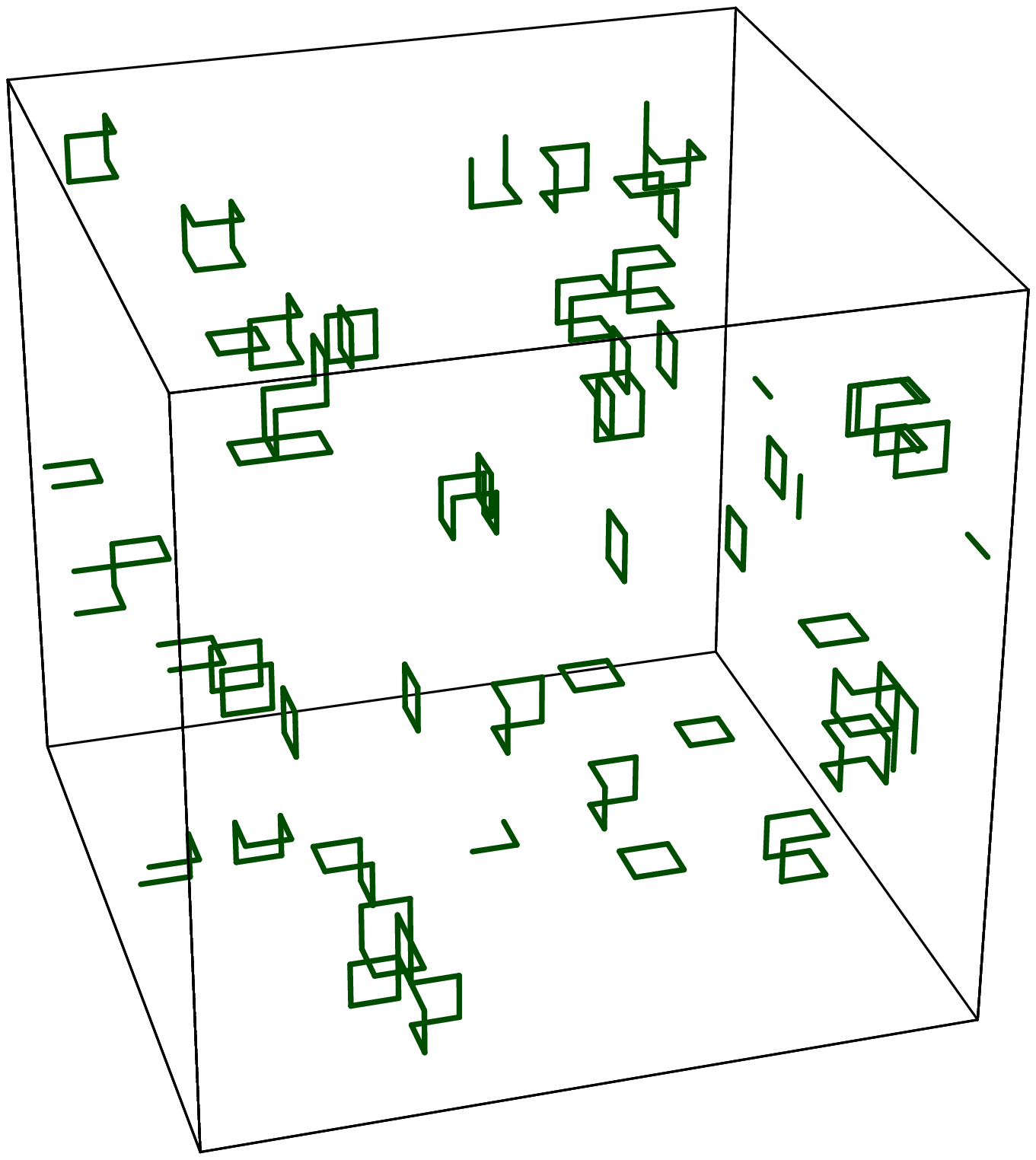}} 
\vspace{0.1cm}
\centerline{(b)}
\vspace{1cm}
\centerline{Fig.~2: The examples of typical abelian diagonal vortex
configurations}
\centerline{in the massive (a) and the massless (b) phases on the $16^3$
lattice.}
\end{figure}

\newpage

\begin{figure}[htb]
\vspace{-6cm}
\centerline{\epsfxsize=.9\textwidth\epsfbox{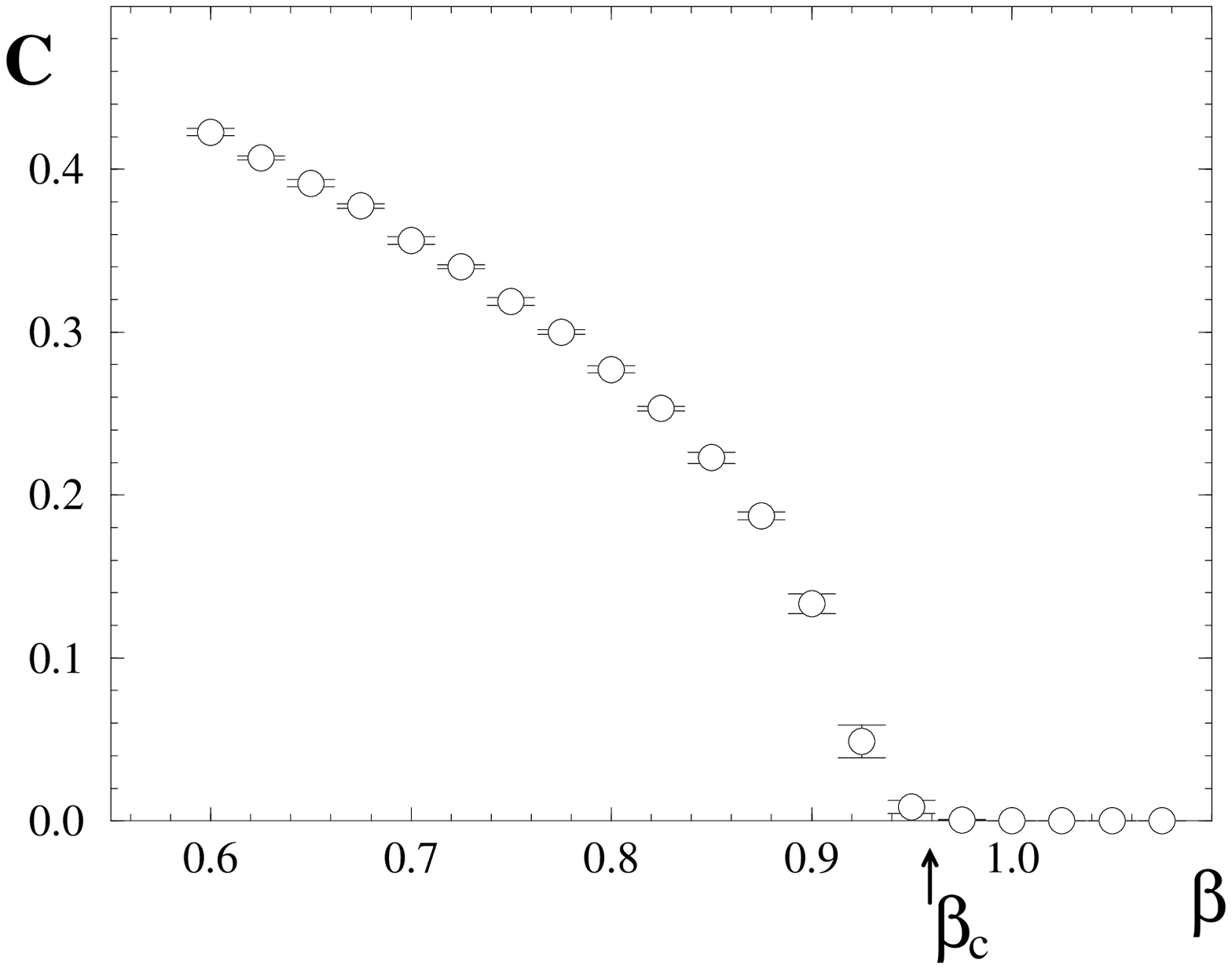}} 
\vspace{0.1cm}
\centerline{Fig.~3: The percolation probability $C$ for the diagonal vortex
trajectories {\it vs.} $\beta$ on the $16^3$ lattice.}
\end{figure}

\end{document}